\documentclass[aps,twocolumn]{revtex4}
\usepackage{epsf}
 
\begin{document}
      \title{Flow equation approach to the pairing problems}

     \author{T.\ Doma\'nski}
\affiliation{
             Institute of Physics, 
	     M.\ Curie Sk\l odowska University, 
             20-031 Lublin, Poland} 
      \date{\today}

\begin{abstract}
We apply the flow equation method for studying the fermion systems 
where pairing interactions can either trigger the BCS instability 
with the symmetry breaking manifested by the off-diagonal order parameter 
or lead to the gaped single particle spectrum without any symmetry 
breaking. We construct the continuous Bogoliubov transformation 
in a scheme resembling the renormalization group procedure. We 
further extend this continuous transformation to a case where 
fermion pairs interact with the boson field. Due to temporal 
quantum fluctuations the single particle excitation spectrum 
develops a gap which is centered around the renormalized 
boson energy. When bosons undergo the Bose Einstein condensation 
this structure evolves into the BCS spectrum.
\end{abstract}  


\maketitle

The flow equation method proposed a couple of years ago by 
Wegner \cite{Wegner-94} and independently by Wilson and 
G\l azek \cite{Wilson-94} turned out to be a very useful 
theoretical tool in the studies of condensed matter and high 
energy physics (for a recent survey of applications see 
for instance \cite{flow-review}). The main idea of this 
technique is to transform the Hamiltonian to a diagonal or 
to a block-diagonal structure using the continuous unitary 
transformation with a flexibly adjusted generating operator. 

A continuous diagonalization can be thought of as a process 
of renormalizing the coupling constants in a way similar to 
the Renormalization Group (RG) approach \cite{Wilson_NRG}. 
Difference between these methods is such that using the flow 
equation procedure one does not integrate out the high energy 
excitations (fast modes). Instead of this, they are renormalized 
in the initial part of the transformation. On the other hand 
the low energy excitations (slow modes) are transformed mainly 
at the very end. In the condensed matter these low energy 
excitations are most relevant for the physical properties. 

In a first part of this paper we show how to construct the 
continuous unitary transformation for the reduced BCS model. 
Exact solution can be obtained of course by various ways, 
in particular via the Bogoliubov-Valatin transformation 
\cite{Bogoliubov}. We rederive here the rigorous solution  
using the flow equation method. This worthwhile to do  
because the pairing instability problem is in the RG studies 
not a trivial issue. Only very recently there appeared
in the literature several concepts based on the functional 
RG \cite{Salmhofer-04,Kopietz-05} discussing how to work out 
the Cooper instability. Since the flow equation method is 
relative to the RG scheme we think it would be instructive 
to follow Wegner {\em et al} \cite{Stelter} and show how 
one can deal with the pairing interactions.

In the second part we generalize this transformation to a case 
of fermions interacting with the boson field. Such situation 
can refer for example to the conduction band electrons coupled 
through the Andreev type scattering to the localized electron 
pairs \cite{Ranninger}. On more general grounds it can also 
describe some boson type modes interacting with the correlated 
electrons \cite{Chubukov_etal}. The other realization of this 
scenario is possible in the ultracold gases where fermion 
atoms interact with the weakly bound molecules giving 
rise to the Feshbach resonance and ultimately leading to 
the resonant superfluidity  \cite{ultracoldatoms}.

\section{The BCS pairing}

In order to have a simple example how the fast and 
slow energy modes are scaled during the continuous 
unitary transformation we consider first the exactly 
solvable bilinear Hamiltonian
\begin{eqnarray}
\hat{H} & = & \sum_{{\bf k},\sigma} \xi_{\bf k} 
\hat{c}_{{\bf k}\sigma}^{\dagger} \hat{c}_{{\bf k}\sigma} 
- \sum_{\bf k} \left( \Delta_{\bf k}  
\hat{c}_{{\bf k}\uparrow}^{\dagger} \hat{c}_{-{\bf k}
\downarrow}^{\dagger} + h.c. \right) 
\label{hamil}
\end{eqnarray}
describing fermions coupled to some pairing field 
$\Delta_{\bf k}$. In (\ref{hamil}) we use the standard 
notation for the creation (annihilation) operator 
$\hat{c}_{{\bf k}\sigma}^{\dagger}$ 
($\hat{c}_{{\bf k}\sigma}$) and energy $\xi_{\bf k}=
\varepsilon_{\bf k}-\mu$ is measured from the chemical
potential $\mu$. Pairing field can be thought of,
for instance, as a mean field approximation to 
the weak pairing potential between electrons  
$\Delta_{\bf k}=-\sum_{\bf q} V_{{\bf k},{\bf q}} 
\langle \hat{c}_{-{\bf q}\downarrow}\hat{c}_{{\bf q}
\uparrow}\rangle$, where $V_{{\bf k},{\bf q}}<0$. 
In general, one can treat $\Delta_{\bf k}$ as 
the boson operator e.g.\ arising from the Hubbard 
Stratonovich transformation for the interacting fermion 
system. It can also correspond to some boson field
responsible for mediating the pairing between electrons.
Coupling to the boson operator will be discussed in 
the next section.

Hamiltonians of the fermion and/or boson systems 
with a bilinear structure (\ref{hamil}) have been 
already considered using the flow equation method 
\cite{Wegner-94, Stelter,Stein,Safonov}. However, 
the authors have focused so far on determination 
of the quasiparticles energies. Here we supplement 
their analysis by calculating the correlation 
functions which contain information about the order 
parameter as well as a characteristics of the 
excitation spectrum. 

It is well known, that the bilinear Hamiltonian (\ref{hamil}) 
can be diagonalized by the (single step) Bogoliubov 
transformation \cite{Bogoliubov}. All physical quantities, 
static and dynamical, can hence be determined exactly. 
In this work we get the same rigorous solution in 
a process of continuous diagonalization of the coupled 
states $|{\bf k},\uparrow>$ and $|-{\bf k},\downarrow>$. 
How fast this can be achieved depends on 
a distance from the Fermi surface $|{\bf k}-{\bf k}_F|$ 
(or on the energy $\xi_{\bf k}$). Simultaneously with
diagonalization there emerge the coherence factors 
$u_{\bf k}$, $v_{\bf k}$. Their final values depend 
on the relative distance of the momentum ${\bf k}$ 
from the Fermi surface.

We construct the continuous unitary transformation 
$\hat{{\cal{U}}}(l)$, where $l$ stands for a formal 
{\em flow parameter} varying between the initial $l=0$ 
value and value $l>0$ such, that transformed 
Hamiltonian $\hat{H}(l)=\hat{{\cal{U}}}(l) \hat{H} 
\hat{{\cal{U}}}^{-1}(l)$ simplifies to the needed 
structure (diagonal, block-diagonal, tridiagonal or 
any other). With an increase of the flow parameter 
$l$ the Hamiltonian evolves according to the general 
flow equation \cite{Wegner-94}
\begin{eqnarray}
\frac{d\hat{H}(l)}{dl} = [ \hat{\eta}(l),\hat{H}(l) ]
\label{general}
\end{eqnarray} 
where $\hat{\eta}(l) \equiv \frac{d\hat{{\cal{U}}}(l)}{dl} 
\hat{{\cal{U}}}^{-1}(l)$. We now require the transformed 
Hamiltonian $\hat{H}(l)$ to preserve its bilinear structure 
(\ref{hamil}) and let only the coupling constants to 
become $l$-dependent (renormalized)
\begin{eqnarray}
\hat{H}(l) & = & \sum_{{\bf k},\sigma} \xi_{\bf k}(l)
\hat{c}_{{\bf k}\sigma}^{\dagger} \hat{c}_{{\bf k}\sigma}
- \sum_{\bf k} \left( \Delta_{\bf k}(l)
\hat{c}_{{\bf k}\uparrow}^{\dagger} \hat{c}_{-{\bf k}
\downarrow}^{\dagger} + h.c. \right) \nonumber \\ 
& + & const(l) \label{renorm}
\end{eqnarray}   
with the initial conditions $\xi_{\bf k}(0)=\xi_{\bf k}$,
$\Delta_{\bf k}(0)=\Delta_{\bf k}$ and $const(0)=0$.
This strategy reminds the scheme of RG approach. It
can be shown \cite{Wegner-94} that, roughly speaking, 
the {\em flow parameter} $l$ corresponds to $\Lambda^{-1}$.

There are several ways for choosing the generating
operator $\hat{\eta}(l)$. In order to drive the Hamiltonian 
to a diagonal structure we adopt the Wegner's proposal 
\cite{Wegner-94} $\hat{\eta}(l)=[ \hat{H}_0(l),\hat{H}(l) ]$, 
where $\hat{H}_0(l)=\sum_{{\bf k},\sigma} \xi_{\bf k}(l) 
\hat{c}_{{\bf k}\sigma}^{\dagger} \hat{c}_{{\bf k}\sigma}$ 
denotes a diagonal part. With this choice the Hamiltonian 
becomes diagonal in the limit $l \rightarrow \infty$. In 
explicit form the generating operator for (\ref{renorm}) 
reads 
\begin{eqnarray}                                                 
\hat{\eta}(l) = - 2 \sum_{\bf k} \xi_{\bf k}(l) \left(
\Delta_{\bf k}(l) \hat{c}_{{\bf k}\uparrow}^{\dagger} 
\hat{c}_{-{\bf k}\downarrow}^{\dagger} -h.c. \right) .
\label{eta}
\end{eqnarray}
Substituted into the equation (\ref{general}) it gives
\begin{eqnarray}
\frac{d\hat{H}(l)}{dl} & = & 4 \sum_{{\bf k},\sigma} 
\xi_{\bf k}(l)|\Delta_{\bf k}(l)|^{2} \left( \hat{c}_{{\bf k}
\sigma}^{\dagger} \hat{c}_{{\bf k}\sigma} -1 \right) \nonumber \\
& - & 4 \sum_{\bf k} (\xi_{\bf k}(l))^{2} \left( \Delta_{\bf k}^{*}(l)
\hat{c}_{{\bf k}\uparrow}^{\dagger} \hat{c}_{-{\bf k}\downarrow}^{\dagger}
+ h.c. \right)  
\label{eq5}
\end{eqnarray}  
which is identical with the following set of flow equations
\begin{eqnarray}
\frac{d\xi_{\bf k}(l)}{dl} & = & 4 \xi_{\bf k}(l) 
|\Delta_{\bf k}(l)|^{2} \label{xi_flow} 
\\
\frac{d\Delta_{\bf k}(l)}{dl}  & = & -4 (\xi_{\bf k}(l))^{2}
\Delta_{\bf k}^{*}(l) \label{Delta_flow} 
\end{eqnarray} 
and $d \; const(l)/dl=-4\sum_{{\bf k},\sigma} \xi_{\bf k}(l)
|\Delta_{\bf k}(l)|^{2}$. 

Formally, solution of the equation (\ref{Delta_flow}) can
be given as
\begin{eqnarray} 
| \Delta_{\bf k}(l)| = |\Delta_{\bf k}| e^{-4\int_{0}^{l} dl'
[\xi_{\bf k}(l')]^2}
\end{eqnarray}
which yields that $\lim_{l \rightarrow \infty} \Delta_{\bf k}(l)=0$ 
for all ${\bf k} \neq {\bf k}_F$. When we multiply equation 
(\ref{xi_flow}) by $\xi_{\bf k}(l)$ and equation (\ref{Delta_flow}) 
by $\Delta_{\bf k}(l)^{*}$ then their sum gives the following 
invariance
\begin{eqnarray} 
\frac{d}{dl} \left\{ (\xi_{\bf k}(l))^{2} + 
|\Delta_{\bf k}(l)|^{2} \right\} = 0 .
\label{invar} 
\end{eqnarray} 
This constraint together with vanishing of $\Delta_{\bf k}(l)$
in the limit $l\rightarrow\infty$ implies the known Bogoliubov 
spectrum
\begin{eqnarray}
\xi_{\bf k}(\infty) = \mbox{sgn}({\xi_{\bf k}}) 
\sqrt{(\xi_{\bf k})^{2} + |\Delta_{\bf k}|^{2} } .
\label{bogol}
\end{eqnarray} 
%
\begin{figure}
\epsfxsize=8cm
\centerline{\epsffile{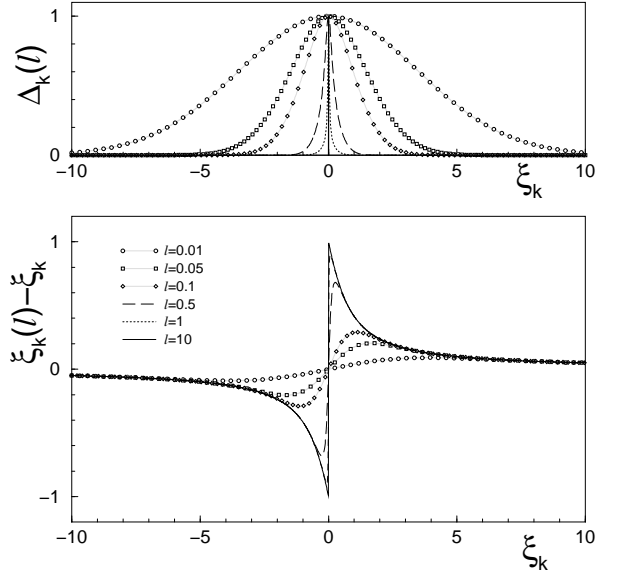}}
\caption{Renormalization of $\Delta_{\bf k}(l)$ and 
$\xi_{\bf k}(l)$ during the {\em flow} for several 
values of $l$ as denoted in the legend. At the 
starting point $l=0$ we used the isotropic coupling 
$\Delta_{\bf k}(0)=\Delta$. Energies are expressed 
in units of $\Delta$ while $l$ in units of $\Delta^{-2}$.}
\label{Fig1}
\end{figure}

In figure \ref{Fig1} we plot the pairing field 
$\Delta_{\bf k}(l)$ and the change (renormalization) of 
fermion energies $\xi_{\bf k}(l)-\xi_{\bf k}$ at several 
stages of the continuous transformation. We notice, that 
fast modes (states distant from the Fermi energy) are 
transformed rather in the first part of the process and 
change of their energies is rather small. The slow modes 
(i.e.\ the low energy excitations) have to be worked on 
much longer. Asymptotically, at $l \rightarrow \infty$, 
the entire spectrum reduces to the Bogoliubov structure 
(\ref{bogol}).

Now we turn attention to the dynamical quantities 
which can be expressed via the normal $\langle \langle 
\hat{c}_{{\bf k}\sigma}; \hat{c}_{{\bf k}\sigma}^{\dagger}
\rangle \rangle_{\omega}$ and the anomalous single particle 
Green's function $\langle \langle \hat{c}_{{\bf k}\uparrow}; 
\hat{c}_{-{\bf k}\downarrow}\rangle \rangle_{\omega}$. We 
introduced the standard Fourier transforms for the retarded 
fermion Green's function $\int d\omega e^{i\omega t} 
\langle \langle \hat{A};\hat{B}\rangle \rangle_{\omega} 
\equiv -i\Theta(t)\langle \hat{A}(t)\hat{B} + \hat{B} 
\hat{A}(t) \rangle$ and time evolution is given  by
$\hat{A}(t)=e^{it\hat{H}}\hat{A}e^{-it\hat{H}}$.

Thermal averaging for some arbitrary observable $\hat{O}$ 
is defined as $\langle \hat{O} \rangle = \mbox{Tr} 
\left\{ e^{-\beta \hat{H}} \hat{O} \right\}/\mbox{Tr} 
\left\{ e^{-\beta \hat{H}}\right\}$, where 
$\beta^{-1}=k_{B}T$. Since the trace is invariant under 
unitary transformations we can write down
\begin{eqnarray}
\mbox{Tr} \left\{  e^{-\beta \hat{H}} \hat{O} \right\} & = &
\mbox{Tr} \left\{ \hat{{\cal{U}}}(l) e^{-\beta \hat{H}} \hat{O} 
\hat{{\cal{U}}}^{-1}(l) \right\}    
\nonumber \\ 
& = & \mbox{Tr} \left\{  e^{-\beta \hat{H}(l)} \hat{O}(l) \right\}
,\label{trace}
\end{eqnarray}
where $\hat{O}(l)=\hat{{\cal{U}}}(l) \hat{O} \hat{{\cal{U}}}^{-1}(l)$. 
It is convenient to compute such trace in the limit $l=\infty$ 
because Hamiltonian becomes then diagonal. However, the price 
which we pay for having the simplified Hamiltonian at $l=\infty$ 
is the additional necessity to transform the observables 
$\hat{O} \rightarrow \hat{O}(l) \rightarrow \hat{O}(\infty)$. 
This is to be done similarly as with the Hamiltonian 
(\ref{general}) using 
\begin{eqnarray}
\frac{d\hat{O}(l)}{dl} = [ \hat{\eta}(l),\hat{O}(l) ] .
\label{O_flow}
\end{eqnarray}   

With the choice of the generating operator $\eta(l)$ 
given by (\ref{eta}) we  find from the commutator 
$[\hat{\eta}(l),\hat{c}_{{\bf k}\uparrow}])$ that 
for $l > 0$ the annihilation operator $\hat{c}
_{{\bf k}\uparrow}$ gets convoluted with 
$\hat{c}_{-{\bf k}\downarrow}^{\dagger}$. It is 
natural to propose the following Ansatz for 
the $l$-dependent operator   
\begin{eqnarray}
\hat{c}_{{\bf k}\uparrow}(l) = u_{\bf k}(l) \hat{c}_{{\bf k}
\uparrow}+v_{\bf k}(l) \hat{c}_{-{\bf k}\downarrow}^{\dagger}. 
\label{Ansatz1}
\end{eqnarray}
Similar analysis for $\hat{c}_{-{\bf k}\downarrow}^{\dagger}$ 
operator leads to 
\begin{eqnarray} 
\hat{c}_{-{\bf k}\downarrow}(l)^{\dagger} =  - v_{\bf k}(l) 
\hat{c}_{{\bf k}\uparrow} + u_{\bf k}(l)
 \hat{c}_{-{\bf k}\downarrow}^{\dagger} 
\label{Ansatz2}
\end{eqnarray} 
with the initial conditions $u_{\bf k}(0)=1$, $v_{\bf k}(0)=0$.
These parametrized equations (\ref{Ansatz1},\ref{Ansatz2}) 
substituted into (\ref{O_flow}) with $\eta(l)$ given in 
(\ref{eta}) lead to the coupled flow equations 
\begin{eqnarray}
\frac{du_{\bf k}(l)}{dl} & = & 2 \xi_{\bf k}(l)
\Delta_{\bf k}(l)  v_{\bf k}(l),
\label{u_flow} \\
\frac{dv_{\bf k}(l)}{dl}  & = & -2 \xi_{\bf k}(l)
\Delta_{\bf k}(l) u_{\bf k}(l) .
\label{v_flow}
\end{eqnarray}  
After straightforward algebra using equations 
(\ref{u_flow},\ref{v_flow}) we obtain the following 
invariance $|u_{\bf k}(l)|^2 + |v_{\bf k}(l)|^2=
\mbox{const}=1$ which assures that the $l$-dependent 
fermion operators (\ref{Ansatz1},\ref{Ansatz2}) 
preserve the anticommutation relations
\begin{eqnarray}
\left\{ \hat{c}_{{\bf k}\sigma}(l),
\hat{c}_{{\bf k'}\sigma'}^{\dagger}(l)\right\} 
= \delta_{{\bf k},{\bf k'}}
\delta_{\sigma,\sigma'}. 
\label{anticommutation}
\end{eqnarray}  

Without a loss of generality we assume $\Delta_{\bf k}(l)$ 
to be real so that the unknown coefficients $u_{\bf k}(l)$, 
$v_{\bf k}(l)$ become real too. To determine their 
asymptotic $l=\infty$ values we can rewrite (\ref{u_flow}) 
as
\begin{eqnarray}
\frac{du_{\bf k}(l)}{v_{\bf k}(l)} = 2 \xi_{\bf k}(l)
\Delta_{\bf k}(l) dl 
\label{eq18}
\end{eqnarray} 
and we further integrate both sides in the limits 
$\int_{l=0}^{l=\infty}$. Using $v_{\bf k}(l)=
\sqrt{1-(u_{\bf k}(l))^{2}}$ we get for the l.h.s.
\begin{eqnarray}
\int_{l=0}^{l=\infty} \frac{du_{\bf k}(l)}
{\sqrt{1-(u_{\bf k}(l))^{2}}}
= \mbox{arcos} \; u_{\bf k}(\infty)
\label{lhs} 
\end{eqnarray} 
due to $u_{\bf k}(0)=1$. Using equation 
(\ref{Delta_flow}) we can replace  $2 \xi_{\bf k}(l)
\Delta_{\bf k}(l)dl$ by $-d\Delta_{\bf k}(l)
/ 2\xi_{\bf k}(l) $ and from the invariance 
(\ref{invar}) we obtain that $\xi_{\bf k}(l)=
\sqrt{\xi_{\bf k}^{2}(\infty)-|\Delta_{\bf k}(l)|^{2}}$.
The r.h.s.\ of equation (\ref{eq18}) gives
after integration
\begin{eqnarray} 
\int_{l=0}^{l=\infty} 2 \xi_{\bf k}(l) 
\Delta_{\bf k}(l) dl & = & - \; \int_{l=0}^{l=\infty}
\frac{d\Delta_{\bf k}(l)}{2\sqrt{\xi_{\bf k}^{2}(\infty)-
|\Delta_{\bf k}(l)|^{2}}} \nonumber \\
& = & \frac{-1}{2} \left( 
\mbox{arcos} \; \frac{|\Delta_{\bf k}|}
{|\xi_{\bf k}(\infty)|} - \frac{\pi}{2} \right)  
\label{rhs} 
\end{eqnarray}
because $\Delta_{\bf k}(\infty)=0$.
Combining the results (\ref{lhs},\ref{rhs}) 
\begin{eqnarray}
2 \; \mbox{arcos} \; u_{\bf k}(\infty) = \frac{\pi}{2} -
\mbox{arcos} \; \frac{|\Delta_{\bf k}|}{|\xi_{\bf k}(\infty)|}
\end{eqnarray} 
we finally determine the $l\!=\!\infty$ factors
\begin{eqnarray} 
u_{\bf k}^{2}(\infty) = \frac{1}{2}
\left[ 1 + \frac{\xi_{\bf k}}{\xi_{\bf k}(\infty)} 
\right] = 1 -  v_{\bf k}^{2}(\infty) . 
\end{eqnarray}    
One can simply show that
$2u_{\bf k}(\infty)v_{\bf k}(\infty)=\Delta_{\bf k}/
|\xi_{\bf k}(\infty)|$.

\begin{figure}
\epsfxsize=7cm
\centerline{\epsffile{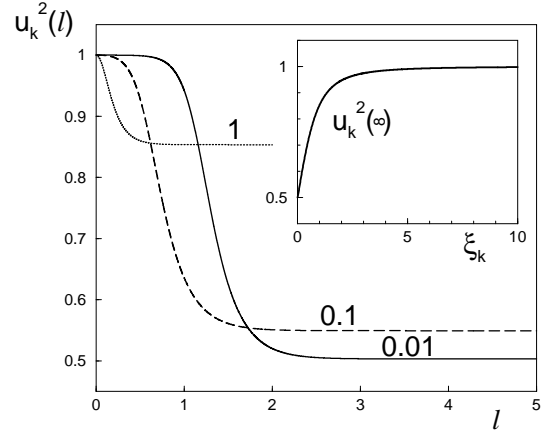}}
\caption{{\em Flow} of the coherence factor 
$u_{\bf k}^{2}(l)$ for three fermion states
of the initial energy $\xi_{\bf k}=0.01$, $0.1$
and $1$ (in units $\Delta$). Inset shows
the effective value of the coherence factor
$u_{\bf k}^{2}(\infty)$ versus $\xi_{\bf k}$.}
\label{Fig2}
\end{figure}

In figure \ref{Fig2} we present the coherence factor 
$u_{\bf k}^{2}(l)$ as a function of the flow parameter 
$l$. Again we notice that for momenta distant from 
the Fermi surface the coherence factors evolve rather 
quickly from the initial value $u_{\bf k}^{2}(0)=1$ 
to the asymptotic values (see the inset). The coherence 
factors of the slow modes establish later on, similarly 
to renormalization of the corresponding excitation 
energies shown in figure \ref{Fig1}. Saturation  
occurs around $l \propto 1/\xi_{\bf k}(\infty)^{2}$.

Since the transformed Hamiltonian $\hat{H}(\infty)$ is 
diagonal we can easily derive the single particle Green's 
functions (and the higher order Green's functions too).
With the parameterizations (\ref{Ansatz1},\ref{Ansatz2})
we obtain 
\begin{eqnarray}
\langle \langle \hat{c}_{{\bf k}\uparrow}; 
\hat{c}_{{\bf k}\uparrow}^{\dagger}\rangle \rangle_{\omega} 
& = &\frac{u_{\bf k}^{2}(\infty)}{\omega-\xi_{\bf k}(\infty)}
+\frac{v_{\bf k}^{2}(\infty)}{\omega+\xi_{\bf k}(\infty)}
\\
\langle \langle \hat{c}_{{\bf k}\uparrow}; 
\hat{c}_{-{\bf k}\downarrow} \rangle \rangle_{\omega}  
& = & \frac{ u_{\bf k}(\infty) v_{\bf k}(\infty) }
{\omega-\xi_{\bf k}(\infty)} - \frac{u_{\bf k}(\infty) 
v_{\bf k}(\infty) } {\omega+\xi_{\bf k}(\infty)} .
\end{eqnarray}
The equal time expectation values defined by 
$\langle \hat{B}\hat{A} \rangle  = - \frac{1}{\pi} 
\int_{-\infty}^{\infty} d\omega f(\omega,T) \mbox{Imag} 
\langle \langle \hat{A}; \hat{B} \rangle \rangle_{
\omega+i0^{+}}$ (where $f(\omega,T)=\left[ e^{\beta 
\omega}+1 \right]^{-1}$ is the Fermi distribution
function) yield the equations for average momentum 
occupancy 
\begin{eqnarray}
\langle  \hat{c}_{{\bf k}\uparrow}^{\dagger} 
\hat{c}_{{\bf k}\uparrow} \rangle  = \frac{1}{2} 
\left[ 1 - \frac{\xi_{\bf k}}{\xi_{\bf k}(\infty)} 
\mbox{tanh}\frac{\xi_{\bf k}(\infty)}{2k_{B}T} \right], 
\label{eq26}
\end{eqnarray} 
and for the off-diagonal order parameter 
\begin{eqnarray}
\langle \hat{c}_{-{\bf k}\downarrow} \hat{c}_{{\bf k}\uparrow} 
\rangle = \frac{\Delta_{\bf k}}{2\xi_{\bf k}(\infty)}
\mbox{tanh}\frac{\xi_{\bf k}(\infty)}{2k_{B}T}.  
\label{eq27}
\end{eqnarray} 
These equations (\ref{eq26},\ref{eq27}) exactly reproduce 
the rigorous solution of the reduced BCS Hamiltonian.

In appendix A we explain how to generalize the present
treatment to account for the scattering of finite momentum 
fermion pairs.

\section{Coupling to the boson mode} 

We apply here the same method to the non-trivial 
problem describing itinerant fermions coupled to 
some dispersionless boson mode 
\begin{eqnarray}
\hat{H}  =  \sum_{{\bf k},\sigma} \xi_{\bf k}
\hat{c}_{{\bf k}\sigma}^{\dagger} \hat{c}_{{\bf k}
\sigma} + \sum_{{\bf k}} \left( g_{\bf k} \; \hat{b} 
\hat{c}_{{\bf k}\uparrow}^{\dagger} \hat{c}_{-{\bf k}
\downarrow}^{\dagger} + h.c. \right) 
+ \Omega \hat{b}^{\dagger} \hat{b}. 
\nonumber \\ \label{Dicke_model}
\end{eqnarray}
Here the boson mode can be regarded for instance as a 
pairing field derived from the Hubbard Stratonovich
transformation in the system of interacting fermions. 
Model  (\ref{Dicke_model}) can describe also the Andreev 
tunneling between $c$ - fermions and some pair reservoir 
denoted by $b$ - particles. Other possibility is to 
think of the case where itinerant fermions (e.g.\ 
conduction band electrons) coexist and interact with 
some localized fermion pairs \cite{Ranninger}. The model 
(\ref{Dicke_model}) is also often applied to describe 
the ultracold fermion atoms coupled to the weakly 
bound molecules effectively leading to resonant 
Feshbach scattering \cite{Timmermans}.

To keep a conserved total number of particles
$\hat{N}_{tot}=\sum_{{\bf k},\sigma} \hat{c}_{{\bf k}
\sigma}^{\dagger} \hat{c}_{{\bf k}\sigma} + 2
\hat{b}^{\dagger} \hat{b}$ we apply the grand 
canonical ensemble. $\Omega$ stands for the boson 
energy measured from $2\mu$ and  $g_{\bf k}$ 
denotes the coupling constant. It can be shown 
that physical properties of this model depend 
only on a magnitude of $g_{\bf k}$, in other 
words all $|g_{\bf k}|e^{i\phi_{\bf k}}$ lead to 
identical results independently of phase 
$\phi_{\bf k}$.

In the mean field treatment of the Hamiltonian 
(\ref{Dicke_model}) one usually introduces the linearization 
$\hat{b} \hat{c}_{{\bf k}\uparrow}^{\dagger}\hat{c}_{-{\bf k}
\downarrow}^{\dagger} \simeq \langle \hat{b} \rangle 
\hat{c}_{{\bf k}\uparrow}^{\dagger}\hat{c}_{-{\bf k}
\downarrow}^{\dagger} + \hat{b} \langle\hat{c}_{{\bf k}
\uparrow}^{\dagger}\hat{c}_{-{\bf k}\downarrow}^{\dagger}
\rangle$ \cite{Ranninger}. Such idea is based on assumption 
that there exists a finite amount of the Bose Einstein (BE) 
condensation of $b$ particles. Hamiltonian (\ref{Dicke_model}) 
simplifies then  to the ordinary BCS problem, where 
$\Delta \equiv -g\langle \hat{b}\rangle$. Let us however 
remark that the BE condensation of infinitely heavy boson 
(characterized by a discrete energy level $\Omega$) cannot 
occur. Our present analysis based on the flow equation 
procedure shows a possible route to go beyond such 
mean field approximation. 
 
We construct the continuous canonical transformation which
decouples $c$ from $b$ particles. This can be achieved only
approximately by choosing the following generating operator 
\begin{eqnarray}
\hat{\eta}(l) & = & \sum_{\bf k} \left[ 2 \xi_{\bf k}(l) 
- \Omega(l) \right] g_{\bf k}(l) \hat{b}
\hat{c}_{{\bf k}\uparrow}^{\dagger}\hat{c}_{-{\bf k}
\downarrow}^{\dagger} - h.c. 
\label{eta_BF}
\end{eqnarray}
Applying (\ref{eta_BF}) to the flow equation 
(\ref{general}) and going through appropriate normal 
ordering we obtain the $l$-dependent Hamiltonian 
\begin{eqnarray}
\hat{H}(l) & \simeq & \sum_{{\bf k},\sigma} \xi_{\bf k}(l)
\hat{c}_{{\bf k}\sigma}^{\dagger} \hat{c}_{{\bf k}
\sigma} \!+\! \sum_{{\bf k}} \left( g_{\bf k}(l) 
\hat{b} \hat{c}_{{\bf k}\uparrow}^{\dagger} 
\hat{c}_{-{\bf k}\downarrow}^{\dagger} + h.c. \right) 
\nonumber \\
& + &  \Omega(l)  \hat{b}^{\dagger} 
\hat{b} + \sum_{{\bf k},{\bf k'}} U_{{\bf k},{\bf k'}}(l)
\hat{c}_{{\bf k}\uparrow}^{\dagger} \hat{c}_{-{\bf k}
\downarrow}^{\dagger}\hat{c}_{-{\bf k'}\downarrow} 
\hat{c}_{{\bf k'}\uparrow}  
\label{transformed_model}
\end{eqnarray}
with $U_{{\bf k},{\bf k'}}(0)=0$. In a straightforward
way we derive the following set of flow equations
\begin{eqnarray}
\frac{d\xi_{\bf k}(l)}{dl} & = & 2 |g_{\bf k}(l)|^{2} 
\left[ 2 \xi_{\bf k}(l) - \Omega(l)  \right]
n^{B} \label{xi_flow_BF}
\\
\frac{d\Omega(l)}{dl} & = & 2 \sum_{\bf k} 
|g_{\bf k}(l)|^{2} \left[ 2 \xi_{\bf k}(l) 
\!-\! \Omega(l) \right] \left( 2 n_{{\bf k} \sigma}^{F} 
\!-\! 1 \right) \label{omega_flow_BF}
\\
\frac{dg_{\bf k}(l)}{dl} & = & - g_{k}(l) \left[ 
2 \xi_{\bf k}(l) - \Omega(l) \right]^{2}  
\label{g_flow_BF}   
\\
\frac{dU_{{\bf k},{\bf k'}}(l)}{dl} & = & -  g_{\bf k} 
g_{\bf k'}^{*}(l) \left[ 2 \xi_{\bf k}(l) - \Omega(l) \right]  
\nonumber \\ & & - g_{\bf k'} g_{\bf k}^{*}(l) 
\left( 2 \xi_{\bf k'}(l) - \Omega(l) \right) .
\label{U_flow_BF}   
\end{eqnarray}
Here $n_{{\bf k}\sigma}^{F}$ denotes the average fermion 
occupancy of the state $|{\bf k},\sigma\rangle$ and 
$n^{B}=\langle \hat{b}^{\dagger}\hat{b}\rangle$. From 
formal solution of the equation (\ref{g_flow_BF}) 
$g_{\bf k}(l)=g_{\bf k}\; \mbox{exp}\{-\int_{0}^{l} dl'
[2\xi_{\bf k}(l')-\Omega(l')]^{2}\}$ we notice that 
$g_{\bf k}(l\!\rightarrow \!\infty)=0$.

In order to close the set of flow equations 
(\ref{xi_flow_BF}-\ref{xi_flow_BF}) we must determine 
the distribution function $n_{{\bf k}\sigma}^{F}$. 
From analysis of the flow equation (\ref{O_flow}) for 
the annihilation and creation fermion operators we 
come to a conclusion that
\begin{eqnarray}
\hat{c}_{{\bf k}\uparrow}(l) & = & u_{\bf k}(l) \;
\hat{c}_{{\bf k}\uparrow} + v_{\bf k}(l) \; \hat{b} \;
\hat{c}_{-{\bf k}\downarrow}^{\dagger}, \label{nr_1} \\
\hat{c}_{-{\bf k}\downarrow}^{\dagger}(l) & = & 
- v_{\bf k}(l) \; \hat{b}^{\dagger} \; \hat{c}_{{\bf k}\uparrow} 
+ u_{\bf k}(l) \; \hat{c}_{-{\bf k}\downarrow}^{\dagger}. 
\label{nr_2}
\end{eqnarray} 
Here the $l$-dependent factors satisfy the following 
flow equations
\begin{eqnarray}
\frac{u_{\bf k}(l)}{dl} & = & v_{\bf k}(l) g_{\bf k}(l) 
\left[ 2 \varepsilon_{\bf k}(l) - \Omega(l) \right]
( n^{B} + n_{-{\bf k}\downarrow}^{F} ) \label{u_BF}\\ 
\frac{v_{\bf k}(l)}{dl} & = & - u_{\bf k}(l) g_{\bf k}(l) 
\left[ 2 \varepsilon_{\bf k}(l) - \Omega(l) \right] . 
\label{v_BF} 
\end{eqnarray} 
with the boundary condition $u_{\bf k}(0)=1$ and 
$v_{\bf k}(0)=0$. For simplicity in (\ref{u_BF},\ref{v_BF}) 
we neglected the terms proportional to $g_{\bf k}(l) 
U_{{\bf k}_1,{\bf k}_2}(l)$ which might eventually 
contribute some higher order corrections of the order 
$\sim g_{\bf k}^{3}$.  Combining (\ref{u_BF}) and 
(\ref{v_BF}) we obtain the invariance
\begin{eqnarray}
|u_{\bf k}(l)|^{2}+( n^{B} + n_{-{\bf k}\downarrow}^{F})
 |v_{\bf k}(l)|^{2} = \mbox{const} = 1 
\label{invar_BF} 
\end{eqnarray} 
which guarantees that operators $\hat{c}_{{\bf k}\sigma}
^{(\dagger)}(l)$ defined in (\ref{nr_1},\ref{nr_2}) 
obey the anticommutation relations.

Various expectation values $\langle \hat{O} \rangle$
are easy to carry out in the limit $l \rightarrow \infty$ 
because fermions are decoupled from the boson field. 
Some delicate problem causes the two-body interaction 
$U_{{\bf k},{\bf k'}}(\infty)$. As will be shown below 
this potential is small, so in the lowest order 
perturbation theory we incorporate its effect via 
the Hartree shift $U_{{\bf k},{\bf k}}(\infty)
n_{{\bf k}\sigma}^{F}$ to fermion energies. This 
weak point of the present analysis can be however 
systematically improved.

\begin{figure}
\centerline{\epsfxsize=7cm\epsffile{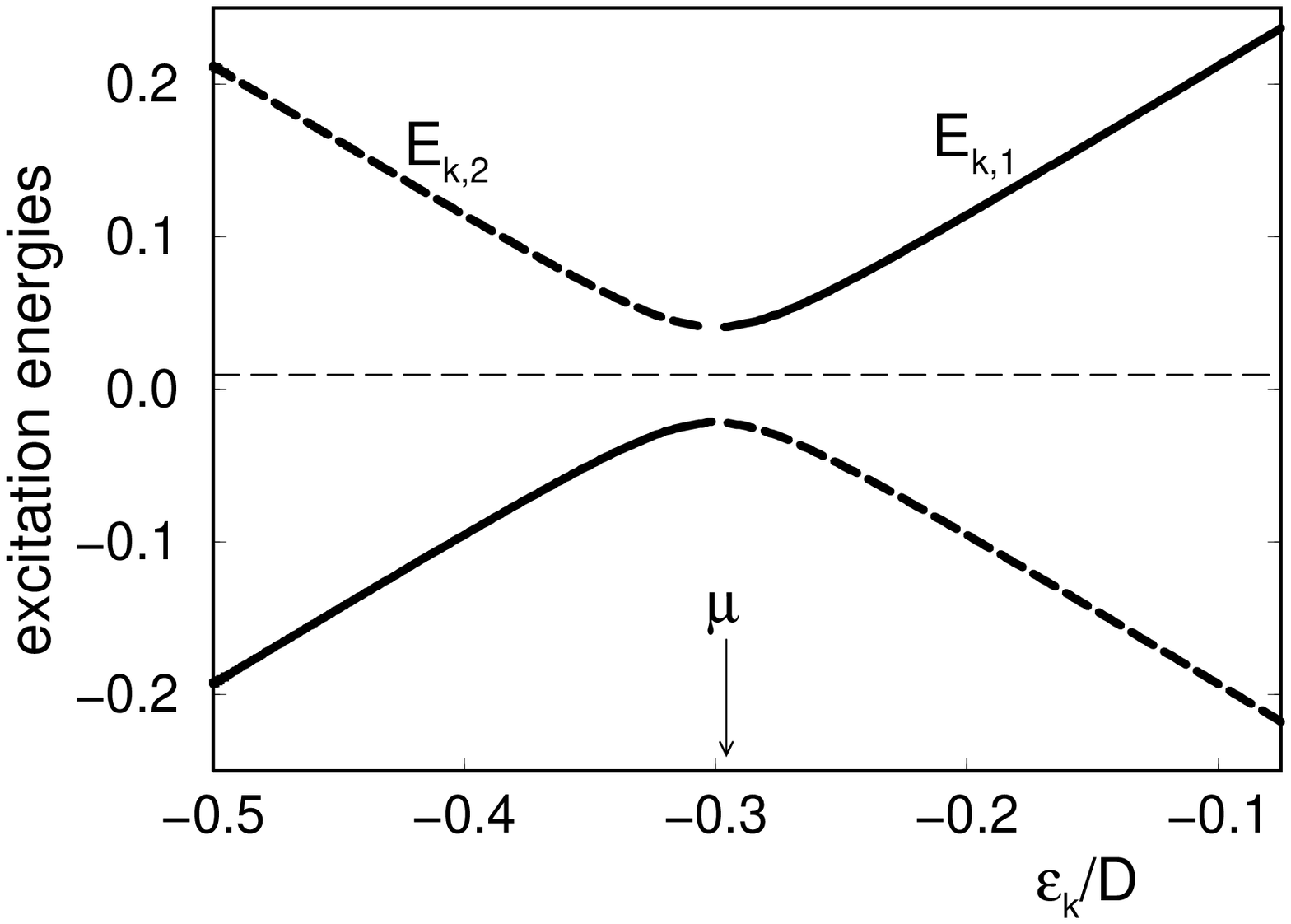}}
\centerline{\epsfxsize=7cm\epsffile{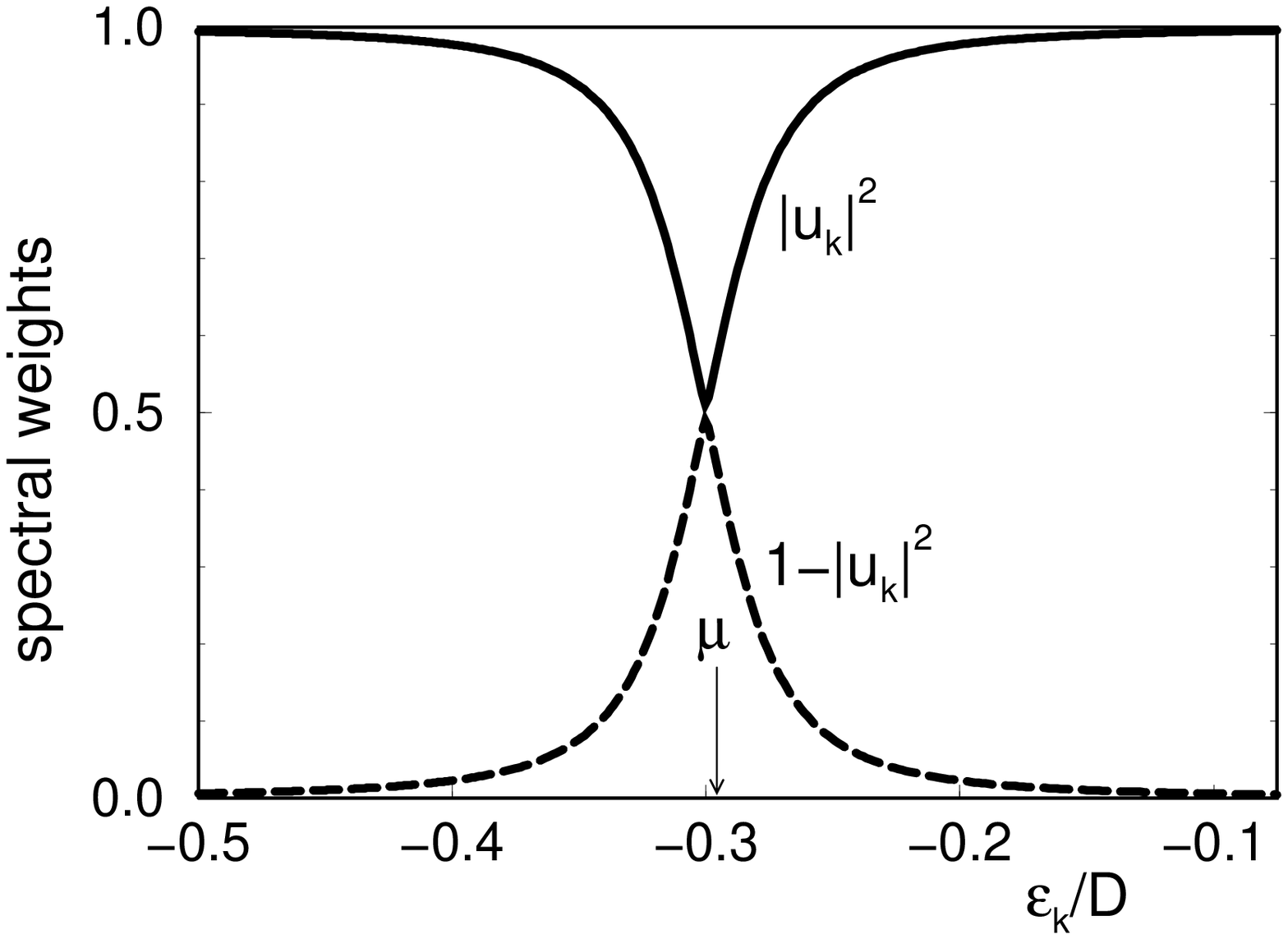}}
\caption{The single particle excitation energies 
(top panel) and their corresponding spectral 
weights (bottom panel). The thin dashed line 
indicates that the excitations $E_{{\bf k},\nu}$
are gaped around the renormalized boson energy
$\frac{1}{2}\Omega(\infty)$. We used
for computations $g_{\bf k}=0.05D$, 
$\Omega/2+\mu=-0.3D$, $k_{B}T=0.015D$.} 
\label{energy}
\end{figure}

Using the parameterization (\ref{nr_1},\ref{nr_2}) 
we find the normal single particle Green's 
function
\begin{eqnarray}
\langle \langle \hat{c}_{{\bf k} \uparrow};
\hat{c}_{{\bf k}\uparrow}^{\dagger} \rangle 
\rangle_{\omega}  =  \frac{| u_{\bf k}(\infty)|^{2}} 
{\omega - E_{{\bf k},1}}  + 
( n^{B} + n_{-{\bf k}\downarrow}^{F} ) \frac{ 
|v_{\bf k}(\infty)|^{2}}{\omega - E_{{\bf k},2}}
\end{eqnarray} 
with quasiparticle energies $E_{{\bf k},1}  = 
\xi_{\bf k}(\infty) + U_{{\bf k},{\bf k}}(\infty) 
n_{{\bf k}\sigma}^{F}$ and $E_{{\bf k},2} = \Omega(\infty) 
\!-\! E_{{\bf k},1}$. The corresponding spectral function 
$A({\bf k},\omega)=-\frac{1}{\pi} \mbox{Imag} \langle 
\langle \hat{c}_{{\bf k}\uparrow};\hat{c}_{{\bf k}
\uparrow}^{\dagger} \rangle \rangle_{\omega + i 0}$ 
satisfies the sum rule $\int_{-\infty}^{\infty} 
d\omega A({\bf k},\omega)=1$ due to invariance 
(\ref{invar_BF}). Finally, the momentum 
distribution function is thus given by
\begin{eqnarray}
n_{{\bf k}\sigma}^{F}  =  \frac{| u_{\bf k}(\infty)|^{2}} 
{e^{\beta E_{{\bf k},1}}+1} \; + \;
( n^{B} + n_{-{\bf k}\downarrow}^{F} ) \;
\frac{| v_{\bf k}(\infty)|^{2}}{e^{\beta E_{{\bf k},2}}+1}.
\label{momentum_distrib}
\end{eqnarray}

We solved numerically the set of coupled flow equations 
(\ref{xi_flow_BF}-\ref{U_flow_BF},\ref{u_BF},\ref{v_BF}) 
and (\ref{momentum_distrib}) using the Runge Kutta algorithm. 
In our calculations we considered the finite fermion band 
$-D/2 \leq \varepsilon_{\bf k} \leq D/2$ and assumed the flat 
density of states. Total particle concentration was fixed 
for $\langle \hat{N}_{tot} \rangle=1$ which corresponds 
to nearly equal populations of fermions and bosons 
$\sum_{\bf k} n_{{\bf k} \sigma}^{F} \sim n^{B}$. 

\begin{figure}
\centerline{\epsfxsize=6cm\epsffile{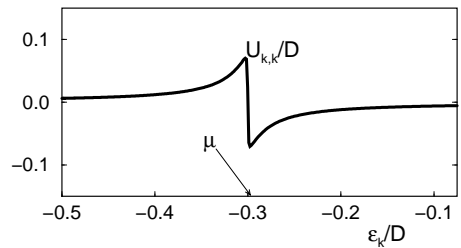}}
\caption{Potential $U_{{\bf k},{\bf k}}(\infty)$ 
of the two-body interaction between fermions for the same 
set of parameters as in figure 3.} 
\label{potencjal}
\end{figure}

In the top panel of figure (\ref{energy}) we show the 
single particle fermion spectrum which turns out to be 
gaped. Two branches of the excitations $E_{{\bf k},\nu}$
are discontinuous around the energy $\Omega(\infty)/2$
(the dashed line). For here considered situation it 
is located slightly above the chemical potential 
$\mu$ and is dependent on temperature. We checked 
that the value of the gap is rather independent of 
$T$. The corresponding spectral weights are shown 
in the bottom panel of figure \ref{energy}. They 
behave similar to the BCS formfactors but we notice 
only a negligible dependence on temperature.
Figure \ref{potencjal} additionally confirms 
that the two-body potential is indeed weak.

We would like to emphasize, that the gaped spectrum shown 
in figure \ref{energy} is not related to any off-diagonal 
order parameter. According to the parameterization (\ref{nr_1},
\ref{nr_2}) we easily determine that the anomalous Green's 
function $\langle \langle\hat{c}_{-{\bf k}\downarrow}
^{\dagger};c_{{\bf k}\uparrow}^{\dagger}\rangle\rangle$ 
is identically zero. The above mentioned structure should 
hence be referred as a pseudogap. The broken symmetry can 
arise if and only if a certain fraction of $b$ particles 
becomes BE condensed \cite{Kostyrko}. There are two 
possibilities for this to occur: 
a) either we assume bosons to be mobile 
   right from the outset of the problem, or
b) we allow for a finite momentum exchange 
   ${\bf q} \neq {\bf 0}$ in the interaction term 
   $\hat{c}_{{\bf k}\uparrow}^{\dagger} 
   \hat{c}_{{\bf q}-{\bf k}\downarrow}^{\dagger}   
   \hat{b_{\bf q}}$.

The second option has been explored in the literature 
by several groups using various many-body techniques 
(for representative list of the references see the review 
paper \cite{Levin-05}). The purpose of our present study 
was to show that the temporal quantum fluctuations alone 
can induce the (pseudo)gap structure - there is no need 
for involving spatial fluctuations. Of course in real 
systems there always exist both, temporal and spatial 
fluctuations. The latter cause that upon lowering 
temperature the (pseudo)gap smoothly evolves into the 
true gap of superconducting state \cite{Domanski-04}.

\section{Conclusions}

By means of the flow equation method \cite{Wegner-94}
we analyzed the bilinear Hamiltonians describing  
fermion systems with pairing interactions. This new 
technique becomes more and more popular \cite{flow-review} 
owing to its conceptual simplicity which allows to go 
beyond the frame of various perturbative approximations  
\cite{Kriel-06}. This method is based on the continuous 
canonical transformations in course of which all the  
parameters are gradually renormalized. In some way 
this can be compared to renormalization of the coupling 
constants within the RG procedure \cite{Wilson_NRG}. 
The high energy excitations are here renormalized mainly 
in the beginning, while the low energy excitations must 
be worked on until the end of transformation.

First part of our paper illustrates how to reproduce
the rigorous solution of the BCS model using a continuous 
version of the Bogoliubov transformation \cite{Bogoliubov}. 
In particular, we show how one can approach the symmetry broken 
problems which usually is not an easy task for the RG methods 
\cite{Salmhofer-04,Kopietz-05}. This scheme can be extended 
for analysis of the two-body interactions where in principle 
various kinds of instabilities can arise. Some results for 
the 2-dimensional Hubbard model have been already reported 
\cite{Wegner_instabilities} with use of the flow equation 
method.

In the second part we focused on the case, where fermion 
pairs are coupled to some single level boson field. Such 
situation can take place when the correlated fermion system 
is affected by bosonic modes such as e.g.\ the pairing 
fluctuations in the systems of reduced dimensionality  
(for instance the HTS cuprates) \cite{PA_Lee-06}. Moreover, 
this sort of physics is recently intensively studied 
for the ultracold fermion atoms where interaction with 
the weakly bound molecules leads to the Feshbach 
resonance which is a driving mechanism for the atomic 
superfluidity \cite{resonant_superfluidity}.

We solved selfconsistently the corresponding set of flow 
equations with a help of the numerics. We showed that 
interactions are responsible for appearance of a gap in 
the single particle fermion spectrum. This gap is centered  
around the boson energy $\frac{1}{2}\Omega(\infty)$ instead 
of the chemical potential. It does not signify any symmetry 
breaking because no order parameter is present in the system. 
The off-diagonal order parameter can eventually appear 
if bosons have a finite mobility (characterized by some  
dispersion $\Omega_{\bf q}$ different from the discrete energy 
$\Omega$ considered here) and if those bosons undergo the BE 
condensation \cite{Kostyrko}. The center of gap moves then 
to the chemical potential as has been explained in the
previous study \cite{Domanski}. Here, we emphasize that 
the single particle gap can appear in a normal state purely 
due to the temporal quantum fluctuations. In realistic 
systems with additional spatial fluctuations such normal 
state (pseudo)gap is expected to evolve smoothly into 
the gap of symmetry broken superconducting state.

\section{Acknowledgment}
Author kindly acknowledges valuable discussions with
professors F.~Wegner, J.~Ranninger and K.I.~Wysoki\'nski.
This work is partly supported by the Polish Scientific
Committee (KBN) through the grant 2P03B06225.

\appendix
\section{} 

The continuous Bogoliubov transformation (discussed 
in section 1) can be extended to the more general 
bilinear Hamiltonian 
\begin{eqnarray}
\hat{H} = \sum_{{\bf k},\sigma} \xi_{\bf k}
\hat{c}_{{\bf k}\sigma}^{\dagger} \hat{c}_{{\bf k}\sigma}
- \sum_{{\bf k},{\bf q}} \left( G_{{\bf k},{\bf q}}
\hat{c}_{{\bf k}\uparrow}^{\dagger} \hat{c}_{{\bf q}-{\bf k}
\downarrow}^{\dagger} + h.c. \right) 
\label{hamil_2}
\end{eqnarray}
which reduces to (\ref{hamil}) if $G_{{\bf k},{\bf q}} 
= \delta_{{\bf q},{\bf 0}} \Delta_{\bf k}$. Besides 
${\bf q}\!=\!{\bf 0}$ this model allows also for 
a scattering of the finite momentum ${\bf q}$ fermion 
pairs. In what follows below we briefly analyze 
the effect of such finite momentum scattering.

The generator from Wegner's proposal 
\begin{eqnarray}
\hat{\eta}^{0}(l) & = & - \sum_{{\bf k},{\bf q}} \left[ 
\xi_{\bf k}(l) + \xi_{{\bf q}-{\bf k}}(l) \right] \left(  
G_{{\bf k},{\bf q}}(l) \hat{c}_{{\bf k}\uparrow}^{\dagger}
\hat{c}_{{\bf q}-{\bf k}\downarrow}^{\dagger}\right) -h.c. 
\nonumber \\  & &
\label{eta_2}
\end{eqnarray}
applied to the flow equation (\ref{general}) induces
the off-diagonal terms $c_{{\bf k}\sigma}^{\dagger} 
c_{{\bf p}\neq{\bf k}\sigma}$. In order to eliminate them from 
the transformed Hamiltonian $H(l)$ we replace 
(\ref{eta_2}) by 
\begin{eqnarray}
\hat{\eta}(l) = \hat{\eta}^{0}(l) + \sum_{{\bf k},{\bf q},\sigma} 
\gamma_{{\bf k},{\bf q},\sigma}(l) \left( 
\hat{c}_{{\bf k} \sigma}^{\dagger} \hat{c}_{{\bf q} \sigma} 
- \hat{c}_{{\bf q}\sigma}^{\dagger} \hat{c}_{{\bf k}\sigma}  
\right). \label{eta_add}
\end{eqnarray}
After simple algebraic calculations we find that 
the off-diagonal terms are exactly canceled if
\begin{eqnarray}
\gamma_{{\bf k},{\bf q},\uparrow}(l) & = &
\sum_{{\bf k'}} \frac{\xi_{\bf k}(l) +
\xi_{{\bf k'}}(l)}{\xi_{\bf k}(l)-\xi_{\bf q}(l)}
G_{{\bf k'},{\bf k'}+{\bf q}-{\bf k}}(l) 
G_{{\bf k},{\bf q}}(l), \nonumber \\ & & \\
\gamma_{{\bf k},{\bf q},\downarrow}(l) & = &
\sum_{{\bf k'}} \frac{\xi_{\bf k}(l) +
\xi_{\bf k'}(l)}{\xi_{\bf k}(l)-\xi_{\bf q}(l)}
G_{{\bf k'},{\bf k}+{\bf k'}}(l) G_{{\bf k'},
{\bf k'}+{\bf q}}(l). \nonumber \\ & &
\end{eqnarray}
The modified generating operator (\ref{eta_add}) has 
a virtue to preserve the initial structure of the Hamiltonian 
$\hat{H}(l) = \sum_{{\bf k},\sigma} \xi_{\bf k}(l)
\hat{c}_{{\bf k}\sigma}^{\dagger} \hat{c}_{{\bf k}\sigma}
- \sum_{{\bf k},{\bf q}} \left( G_{{\bf k},{\bf q}}(l)
\hat{c}_{{\bf k}\uparrow}^{\dagger} \hat{c}_{{\bf q}-{\bf k}
\downarrow}^{\dagger} + h.c. \right)  + const(l)$.
The corresponding set of flow equations is 
\begin{eqnarray}
&& \frac{d\xi_{\bf k}(l)}{dl} = 2 \sum_{\bf q} \left(
\xi_{\bf k}(l) + \xi_{{\bf q}-{\bf k}}(l)  \right)
|G_{{\bf k},{\bf q}}(l)|^{2}, \label{xi_flow_2}
\\
&& \frac{dG_{{\bf k},{\bf q}}(l)}{dl}  = - \left( 
\xi_{\bf k}(l) + \xi_{{\bf q}-{\bf k}}(l)  \right)^{2}  
G_{{\bf k},{\bf q}}^{*}(l) 
\label{Delta_flow_2}  \\ 
&& + 2 \sum_{{\bf k'},\sigma} 
\left( \gamma_{{\bf k},{\bf k'},\sigma}(l) 
- \gamma_{{\bf k'},{\bf k},\sigma}(l)
\right) G_{{\bf k'},{\bf q}-{\bf k}+{\bf k'}}(l),
\nonumber \\ 
&&\frac{ d \; const(l)}{dl} = 
- 2 \sum_{{\bf k},{\bf q},\sigma} ( \xi_{\bf k}(l) 
+ \xi_{{\bf q}-{\bf k}}(l)  )|G_{{\bf k},{\bf q}}(l)|^{2}.
\nonumber \\ & & \label{const_flow_2} 
\end{eqnarray}

We were not able to solve analytically the equations 
(\ref{xi_flow_2}-\ref{const_flow_2}) therefore we 
explored them numerically assuming the tight binding 
dispersion $\xi_{\bf k} = -2t \; \mbox{cos} (k_{x}a)$ 
and using the pairing potential  $G_{{\bf k},{\bf q}}$
in a Lorentian form
\begin{eqnarray}
G_{{\bf k},{\bf q}} = {\cal{N}}(n) \; \frac{\Delta }
{n \left[ q_{x} a \right]^{2} + 1} \;.
\label{lorenz}
\end{eqnarray}
The normalization factor ${\cal{N}}(n)$ was taken such
that $\sum_{\bf q} G_{{\bf k},{\bf q}}=\Delta$ and we 
set $\Delta/4t=0.01$. 

Flow of the operators $c_{{\bf k}\sigma}^{(\dagger)}(l)$  
takes here the form
\begin{eqnarray}
\hat{c}_{{\bf k}\uparrow}(l) & = & 
\sum_{{\bf q}} u_{{\bf k},{\bf q}}(l) 
\hat{c}_{{\bf q}+{\bf k} \uparrow} 
+ \sum_{{\bf q}} v_{{\bf k},{\bf q}}(l) 
\hat{c}_{{\bf q}-{\bf k} \downarrow}^{\dagger} \;, 
\label{Ansatz1_2} \\ 
\hat{c}_{-{\bf k}\downarrow}^{\dagger}(l) & = &  
- \sum_{{\bf q}} v_{{\bf k},{\bf q}}(l) 
\hat{c}_{{\bf q}+{\bf k}\uparrow}
+ \sum_{{\bf q}} u_{{\bf k},{\bf q}}(l)
\hat{c}_{{\bf q}-{\bf k}\downarrow}^{\dagger} 
\label{Ansatz2_2}
\end{eqnarray}
which generalizes the previous equations  (\ref{Ansatz1},
\ref{Ansatz2})  due to finite momentum scattering.
Substituting (\ref{Ansatz1_2},\ref{Ansatz2_2}) into the 
flow equation (\ref{O_flow}) we obtain 
\begin{eqnarray}
&&\frac{du_{{\bf k},{\bf q}}(l)}{dl} =  \left[
\xi_{\bf k}(l) + \xi_{{\bf q}-{\bf k}}(l) \right]
G_{{\bf k},{\bf q}}(l) v_{{\bf k},{\bf 0}}(l) 
\nonumber \\ && +
\sum_{{\bf p}\neq{\bf k}} \left[
\xi_{{\bf k}-{\bf q}}(l) + \xi_{{\bf q}+{\bf p}}(l) \right]
G_{{\bf q}-{\bf k},2{\bf q}-{\bf k}+{\bf p}}(l)
v_{{\bf k},{\bf q}}(l) ,
\nonumber \\ && \label{uII_flow} \\
&& \frac{dv_{{\bf k},{\bf q}}(l)}{dl} =  - \left[
\xi_{\bf k}(l) + \xi_{{\bf q}-{\bf k}}(l)  \right]
G_{{\bf k},{\bf q}}(l) u_{{\bf k},{\bf 0}}(l) 
\nonumber \\  && -
\sum_{{\bf p}\neq{\bf k}} \left[ 
\xi_{{\bf k}-{\bf q}}(l) + \xi_{{\bf q}+{\bf p}}(l) \right]
G_{{\bf q}-{\bf k},2{\bf q}-{\bf k}+{\bf p}}(l)
u_{{\bf k},{\bf q}}(l) .
\nonumber \\ \label{vII_flow} 
\end{eqnarray}
From these equations (\ref{uII_flow}-\ref{vII_flow}) 
we derive the invariance
\begin{eqnarray}
\sum_{\bf q} |u_{{\bf k},{\bf q}}(l)|^{2} +
\sum_{\bf q} |v_{{\bf k},{\bf q}}(l)|^{2} = 1
\label{conserv}
\end{eqnarray}
which again leads to the anticommutation relations
(\ref{anticommutation}).

With use of the parameterization (\ref{Ansatz1_2},\ref{Ansatz2_2}) 
we can now determine the single particle Green's functions. Let 
us first consider the diagonal part (in the Nambu representation). 
The spectral function $A({\bf k},\omega) = - \pi^{-1} {\rm Imag} 
\langle \langle \hat{c}_{{\bf k}\uparrow}; \hat{c}_{{\bf k}
\uparrow}^{\dagger}\rangle \rangle_{\omega+i 0^{+}}$ turns out 
to consist of two contributions
\begin{eqnarray}
A({\bf k},\omega) = A_{coh}({\bf k},\omega) +
A_{inc}({\bf k},\omega) .
\end{eqnarray}
The coherent part 
\begin{eqnarray}
A_{coh}({\bf k},\omega) & = & |u_{{\bf k},{\bf 0}}(\infty)|^{2}
\; \delta \left[ \omega-\xi_{\bf k}(\infty) \right] 
\nonumber \\ & + &
 |v_{{\bf k},{\bf 0}}(\infty)|^{2}
\; \delta \left[ \omega+\xi_{-\bf k}(\infty) \right] 
\end{eqnarray}
describes the long-lived quasiparticle modes of energies
$ \pm \xi_{\bf k}(\infty)$ similar to the Bogoliubov modes 
(\ref{bogol}) discussed previously for $G_{{\bf k},{\bf q}} 
= \Delta_{\bf k} \delta_{{\bf q},{\bf 0}}$. The remaining 
incoherent part 
\begin{eqnarray}
A_{inc}({\bf k},\omega) & = & \sum_{{\bf q}\neq{\bf 0}}  
|u_{{\bf k},{\bf q}}(\infty)|^{2}
\; \delta \left[ \omega-\xi_{{\bf q}+{\bf k}}(\infty) 
\right] 
\nonumber \\ & + &
\sum_{{\bf q}\neq{\bf 0}}  |v_{{\bf k},{\bf q}}(\infty)|^{2}
\; \delta \left[ \omega+\xi_{{\bf q}-\bf k}(\infty) \right].
\label{diag_spectr_funct}
\end{eqnarray}
corresponds to the background spectrum which is spread over 
the large energy regime. Due to invariance (\ref{conserv}) 
the total spectral weight of the coherent and incoherent 
parts satisfies the sum rule $\int_{-\infty}^{\infty} 
d\omega A({\bf k},\omega) = 1$.

\begin{figure}
\epsfxsize=7cm
\centerline{\epsffile{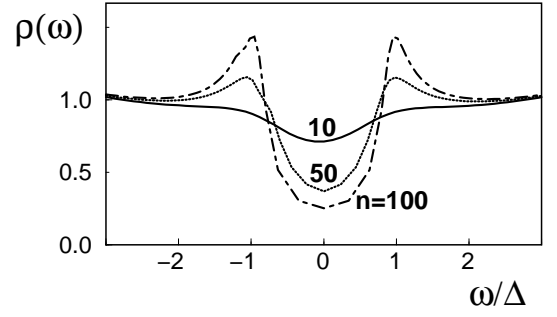}}
\caption{The local density of states $\rho(\omega)=\sum_{\bf k} 
A({\bf k},\omega)$ of fermions coupled to the potential 
$G_{{\bf k},{\bf q}}$ given in (\ref{lorenz}) with $n=10$, 
$50$ and $100$. The density is normalized to $\rho_{0}(\omega)$ 
of the nointeracting sytem.}
\label{Fig5}
\end{figure}

Figure \ref{Fig5} presents the local density of states
calculated by us with use of the definition $\rho(\omega)
=\sum_{\bf k} A({\bf k},\omega)$. We notice that the 
excitation spectrum is not longer truly gaped. There 
occurs only a partial suppression of the fermion states 
around the energy ($\omega=0$). This property should be 
assigned to the scattering of finite momentum fermion  
pairs on the potential $G_{{\bf k},{\bf q}}$ introduced 
ad hoc in (\ref{hamil_2}). Physically this means that 
only a certain fraction of the fermion states is 
expelled from a vicinity of the Fermi surface.

%
\begin{figure}
\epsfxsize=8.5cm
\centerline{\epsffile{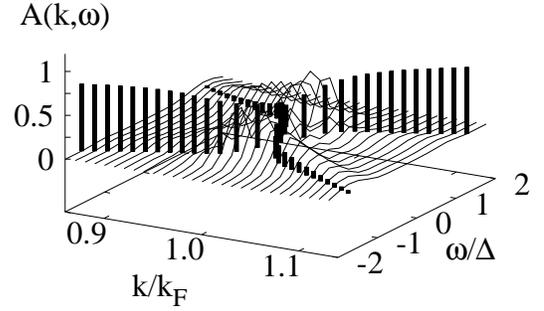}}
\caption{The single particle spectral function $A({\bf k},\omega)$ 
obtained for the potential (\ref{lorenz}) with $n=100$. The thick
vertical lines represent the spectral weights $|u_{{\bf k},{\bf 0}}
(\infty)|^{2}$ and $|v_{{\bf k},{\bf 0}}(\infty)|^{2}$ of the 
long-lived quasiparticles $\pm \xi_{\bf k}(\infty)$. The thin 
solid lines show $\omega$ dependence of the incoherent background 
$A_{inc}({\bf k},\omega)$.}
\label{Fig6}
\end{figure}

Figure (\ref{Fig6}) shows both parts of the spectral function
for the pairing potential (\ref{lorenz}) with $n\!=\!100$, which
yields the strongest suppression of the low lying fermion states 
ploted in the figure \ref{Fig5}. Two modes of the coherent part 
have similar spectral weights as the BCS coherence factors. 
However, in distinction from the standard BCS solution, 
the corresponding quasiparticle dispersion $\pm \xi_{\bf k}
(\infty)$ is in this case gapless. This is evidently caused
by a coupling to the finite momentum fermion pairs. As regards
the incoherent background it builds up mainly near the Fermi 
surface ${\bf k}_{F}$. Such damped fermion states are located 
around the quasiparticle energies $\pm \xi_{\bf k}(\infty)$. 
We estimated that for ${\bf k}={\bf k}_{F}$ the incoherent 
states contribute nearly 30 percent of the total spectral 
weight.

The off-diagonal single particle Green's function 
$\langle \langle \hat{c}_{{\bf k}\uparrow}; 
\hat{c}_{-{\bf p}\downarrow} \rangle \rangle_{\omega}$ 
has a similar structure to (\ref{diag_spectr_funct}).
Skipping the unnecessary technical details we show 
the expression for expectation value of the order 
parameter
\begin{eqnarray}
\langle \hat{c}_{-{\bf p}\downarrow} 
\hat{c}_{{\bf k}\uparrow} \rangle & = & - \frac{1}{2} 
\sum_{{\bf q}} \left\{ \frac{u_{{\bf k},{\bf q}
+{\bf p}-{\bf k}}(\infty) \; v_{{\bf p},{\bf q}}(\infty)} 
{\mbox{exp} \left[ \beta \xi_{{\bf q}+{\bf p}}(\infty)
\right] + 1 } \right.
\nonumber \\
& & -  \left. 
\frac{ u_{{\bf p},{\bf q}}(\infty) \; 
v_{{\bf k},{\bf q}-{\bf p}+{\bf k}}(\infty)} 
{\mbox{exp}\left[ -\beta \xi_{{\bf q}-{\bf p}}(\infty) 
\right] + 1} \right\} .
\end{eqnarray}
We investigated numerically the ${\bf q}$-dependence of the 
order parameter $\langle \hat{c}_{{\bf q}-{\bf k}\downarrow} 
\hat{c}_{{\bf k}\uparrow} \rangle$ and we noticed its rather 
fast decrease against $|{\bf q}|$. For $n=10$, $50$ and 
$100$ we obtained that magnitude of the order parameter at 
$|{\bf q}|=0.2\pi/a$ is circa 100 times smaller than $\langle 
\hat{c}_{-{\bf k}\downarrow} \hat{c}_{{\bf k}\uparrow} \rangle$. 
This indicates that finite momentum fermion pairs are less 
favored in the system. Moreover, the finite momentum fermion
pairs are damped entities because they arise from the incoherent
part of the off-diagonal spectral function existing in 
a broad energy interval $\omega$ for momenta $\sim {\bf k}_{F}$.

\end{document}